
\documentstyle[11pt]{article}
\input pream1
\include{matex}

\def\gsim{\;
\raise0.3ex\hbox{$>$\kern-0.75em\raise-1.1ex\hbox{$\sim$}}\;
}
\def\lsim{\;
\raise0.3ex\hbox{$<$\kern-0.75em\raise-1.1ex\hbox{$\sim$}}\;
}

\thispagestyle{empty}
\begin{titlepage}
\begin{center}
\rightline{\hfill FTUV/94-08}
\rightline{\hfill IFIC/94-05}
\rightline{February, 1994}
\vskip 0.3cm
\LARGE
{\bf SO(10) grand unification model for
degenerate neutrino masses}
\vskip 1cm
\end{center}
\normalsize
\vskip1cm
\begin{center}
{\bf A. Ioannissyan}
\footnote{
On leave from {\em Yerevan Physics Institute,
Armenia}}
{\bf and J.W.F. Valle}
\footnote{E-mail VALLE at vm.ci.uv.es or 16444::VALLE}\\
\end{center}
\begin{center}
\baselineskip=13pt
{\it Instituto de F\'{\i}sica Corpuscular - C.S.I.C.\\
Departament de F\'{\i}sica Te\`orica, Universitat de Val\`encia\\}
\baselineskip=12pt
{\it 46100 Burjassot, Val\`encia, SPAIN         }\\
\vglue 0.8cm
\end{center}

\begin{abstract}

We propose an SO(10) scheme where \neu masses
can simultaneously explain the solar and atmospheric neutrino
deficits, together with a hot dark matter component. In our
scheme the $\nu_e$, $\nu_\mu$, and $\nu_\tau$ are approximately
degenerate with a mass of about 2 eV, which can lead to an
observable neutrinoless double beta decay rate. The model is based
on a realization of the seesaw mechanism in which
the main contribution to the light \neu masses is universal,
due to a suitable SU(2) horizontal symmetry, while
the splittings between \ne and \nm explain the solar
\neu deficit and that between \nm and \nt explain the
atmospheric \neu anomaly.

\end{abstract}

\vfill
\end{titlepage}

\newpage

\vfill\eject

\section{Introduction}

Up to now the only hints in favour of massive \neus
come from astrophysics and cosmology. These involve
four different measurements of a reduced solar neutrino
flux \cite{granadasol}, two or three measurements of a lower
flux of \nm neutrinos relative to \ne neutrinos
produced in the atmosphere \cite{atm}, and the indications
for the need for a neutrino component in the dark
matter of the universe, inferred from the comparison
of recent COBE data \cite{cobe} with data on the
amplitude of primordial fluctuations, e.g. from IRAS,
on smaller distance scales.

While particle physics strongly suggests that
\neus are massive, it unfortunately is unable
to specify the scale that should characterize
the masses of the neutrinos.
Therefore the above solar and atmospheric
\neu deficits plus the indications for hot dark matter in the
universe constitute our only clues into the pattern of
\neu masses.

It was first noted that the simplest way to
reconcile these observations require the
existence of four light neutrinos, one of which
must be sterile \cite{DARK92,DARK92B,DARK92C,DARK92D},
in order to comply with the LEP measurements of the
invisible Z decay width.  The possible patterns of
\neu masses and mixing have already been discussed
and theoretically implemented.

However, another pattern with only three light \neus
is possible if they are almost degenerate in mass \cite{caldwell,joshi}.
This possibility has not attracted as much attention because,
in the simplest seesaw model, the \neu masses are expected
to be in the ratios \mne : \mnm : \mnt = $m_u^2$ : $m_c^2$ : $m_t^2$
and therefore degeneracy is quite unlikely.

It is well known, however, that in the most general seesaw model
there is an additional contribution to the light \neu masses, involving
the vacuum expectation value (VEV) of a triplet scalar boson, as
introduced in ref. \cite{2227}. Moreover, it has been showed
that in left-right symmetric implementations of the seesaw
mechanism such a nonvanishing triplet VEV is always induced \cite{LR}.

In this letter we propose an SO(10) model in which
the main contribution to the effective light \neu
masses arises from such an induced triplet VEV and their
splittings are generated by the usual so-called seesaw
contributions which scale with the quark masses as above.
The model reconciles all three hints for \neu masses and
is consistent with cosmological as well as particle physics
constraints, such as those from the proton decay limits
and the measurements of the electroweak mixing parameter
$\sin^2 2 \theta_W$ and the strong coupling constant $\alpha_s$
derived from LEP and other data. Moreover, it leads to
an observable neutrinoless double beta decay
($\beta\beta_{0\nu}$) rate.

The enhanced $\beta\beta_{0\nu}$ decay rate is rather
intriguing in view of the present experimental situation.
Indeed, as more data from the Heidelberg-Moscow experiment
accumulates the limits do not seem to improve, and may be
interpreted as a 2$\sigma$ effect \cite{Fiorini}. With
current matrix element calculations, this would correspond
to an effective Majorana mass of about 2 eV, of the same
magnitude as found in our scheme.

\section{Preliminaries}

Before describing our SO(10) model, we will first review the
phenomenologically required parameter values of our proposed
solution to the solar, atmospheric and dark matter \neu
mass hints involving almost degenerate Majorana neutrino
masses.

For our proposed solution the solar neutrino deficit
is understood in terms of small-angle (non adiabatic)
\ne to \nm matter enhanced MSW conversions \cite{MSW},
and this requires the following \ne to \nm squared
mass difference and mixing angle \cite{MSWPLOT},
\beq
\label{sol0}
\Delta m^2_{e \mu} \sim 6 \times10^{-6} \rm{eV}^2, \: \
	\sin^22 \theta_{e \mu}\sim 7 \times10^{-3}, \\
\eeq
An apparent decrease in the expected flux of atmospheric
$\nu_\mu$'s relative to $\nu_e$'s arising from the decays
of $\pi$'s and $K$'s produced in the atmosphere, and from
the subsequent secondary muon decays has been observed
in three underground experiments, Kamiokande, IMB
and possibly Soudan2 \cite{atm}. This atmospheric
neutrino deficit can be ascribed to \neu oscillations.
Combining these experimental results with observations
of upward going muons made by Kamiokande, IMB and Baksan,
and with the negative Frejus and NUSEX results \cite{up}
leads to the following range of neutrino oscillation
parameters \cite{atmsasso}
\beq
\label{atm0}
\Delta m^2_{\mu \tau} \approx 0.005 \: - \: 0.5\ \rm{eV}^2,\
\sin^22\theta_{\mu \tau} \approx 0.5
\eeq
Finally, there has been increasing evidence that more than
90\% of the mass in the universe is detectable only by its
gravitational effects.  Recent observations of
the large scale structure of the universe by the COBE
satellite, when taken together with observations of the
amplitude of primordial fluctuations on smaller distance
scales, indicate that this dark matter
is likely to be a mixture of $\sim30$\% of hot dark matter
(particles which were relativistic at the time of freeze-out
from equilibrium in the early universe) and $\sim60$\% of
cold dark matter (particles which were non-relativistic at
decoupling) \cite{cobe,cobe2}. Massive neutrinos
provide the most plausible hot dark matter candidate.
Choosing the total mass
$m_{\nu}=93h^2F_\nu \Omega_{total} = 7$ eV and $h=0.5$
(the Hubble constant in units of 100 km/s/Mpc) one finds that
the fraction of the hot dark matter contributed by \neus
is $F_\nu=0.3$ for $\Omega_{total}=1$ (the ratio of the total
density of the universe to the closure density).  In our scheme
this dark matter is shared amongst the three types of light
neutrinos \ne, \nm and \nt.
It is interesting to note that such a multi component hot dark
matter scenario seems to provide a better fit to the universe
structure than a single $\sim7$ eV neutrino.

In order to account for the above hints for massive
\neus coming from solar and atmospheric neutrino deficits
as well as hot dark matter we use the most economic
pattern of \neu masses consisting of three almost
degenerate neutrinos \ne, \nm and \nt. We focus on
models of the seesaw type, where there exist some
heavy \321 singlet \rh \neus at some large
mass scale $M_R$. The effective mass matrix for
the light \neus may be written as
\beq
\label{mtot}
m_{ij} - {(D M_R^{-1} D^T)}_{ij}
\eeq
Since we wish the three light \neus to be
closely degenerate, we assume that the
the leading contribution is the first, and
that it is proportional to the identity matrix
\beq
\label{m0}
m_{ij} = m \delta_{ij}
\eeq
The corresponding degeneracy in the light \neu spectrum
is lifted since the horizontal symmetry is broken by the
quark masses. This happens through the usual seesaw
corrections that arise from the second term, whose
contribution we assume to be very small.

Assuming that CP is conserved one can write the
charged current leptonic mixing matrix as a product
of three rotations
$K_L = \omega_{(e\mu)} \omega_{(\mu\tau)} \omega_{(e\tau)}$,
in the corresponding (12) , (23) and (13) planes \cite{2227},
times a diagonal phase matrix which is the square root of
\neu CP signs $\eta_i$ \cite{CPS},
\beq
(K_L)_{ai} = K_{ai} \eta_i^{1/2}
\eeq
where
\beq
\label{K}
K = 	\left(\begin{array}{ccccc}
	 C_{12} C_{13}	+ S_{12} S_{13} S_{23} &
	 - S_{12} C_{23} &
        S_{12} C_{13} S_{23} - C_{12} S_{13} \\
	S_{12} C_{13} - C_{12} S_{13} S_{23} &
	C_{12} C_{23} &
       - (S_{12} S_{13} + C_{12} C_{13} S_{23})\\
	S_{13} C_{23} &
        S_{23} &
 	C_{13} C_{23}
\end{array}\right)
\eeq
The required mass parameters that characterize the solar
\neu conversions and atmospheric \neu oscillations
are given by
\beq
\label{sol}
\Delta m^2_{e \mu} \simeq 2 m \Delta m_{e \mu}
\eeq
\beq
\label{atm}
\Delta m^2_{\mu \tau} \simeq 2 m \Delta m_{\mu\tau}
\eeq
where
\beq
\label{2}
m \simeq 2 \rm{eV}
\eeq
and the splittings are determined from \eq{sol0} and
\eq{atm0} to lie in the ranges from $1.5 \times 10^{-6}$ eV,
and $10^{-3}$ to $10^{-1}$ eV, respectively. Moreover,
the mixing angles 12 and 23 are restricted by \eq{sol0} and
\eq{atm0} while 13 is a free parameter.

Finally we note that \cite{joshi}
\beq
\label{ratio}
\frac{\Delta m^2_{e \mu}}{\Delta m^2_{\mu \tau}} =
\O(\frac{m_c^2}{m_t^2} )
\eeq
where $m_c$ and $m_t$ denote the charm and top quark
masses. This relation is the basis of our suggestion of
an SO(10) seesaw origin for the solar and atmospheric
\neu mass splittings. On the other hand the mixings
are given as in \eq{sol0} and \eq{atm0}.

\section{SO(10) model for degenerate Majorana neutrino masses}

In order to account for the required approximate degeneracy
in the \neu spectrum we invoke the existence of a horizontal
symmetry, $G_H$, chosen in such a way that the effective triplet
VEV giving rise to the first term in \eq{mtot} is a $G_H$ singlet,
\eq{m0}.

The particle content of the model is given in tables 1
and 2. It contains only the known fermions which are
assigned to the usual 16-dimensional spinor
representation of SO(10). The three families of such
fermions form a triplet of the global $G_H$=SU(2)$_H$
symmetry.

As for the Higgs sector, the SO(10) symmetry
is assumed to break down at the GUT scale, $M_X$, to the
left-right symmetric group $SU(2)_L \ot SU(2)_R \ot SU(4) \ot P$
via a nonzero VEV of the singlet Higgs in the 54 (see table 1).
It is important to note that the discrete parity
symmetry P implies that the mass of the $SU(2)_L$
and $SU(2)_R$ triplet higgs bosons are of the same
order, which we choose to be determined by $v_R$.
Below we will show that $v_R$ can be determined
to be $\sim 10^{14}$ GeV, a value which fits
well with \neu mass estimated from the large
scale dark matter observations, $m$ = 2 eV.
The SU(4) in the intermediate left-right symmetric
\gau symmetry group is further broken at a scale
$\sim v_R$ via nonzero VEVS of the \321 singlet
Higgs fields $\Delta_R$ in the (10,1,3) of the 126
and the corresponding (15,1,1) of the 45 (see table 1).

Let us first consider how we can get the VEV seesaw
relation for left and right higgs triplets in the 126.
The relevant part of the higgs potential contains
\footnote{Other terms arising from 10's or other possible
representations like 120 or 126 do not change this argument}
\bea
	M^2_{\Delta_L} 126 \: \overline{126}
	+ a_1 126 \: \overline{126} \: 126 \: \overline{126}
	+ a_2 10 \: 10 \: 126 \: 126
	+ a_3 10 \: 10 \: \overline{126} \: \overline{126}
	+ h.c.
\eea
The \321 \gau symmetry is broken by the doublet VEVS in the
10 and 126. After substituting the values of the
corresponding effective $v_u$ and $v_d$ VEVS one finds
\bea
	M^2_{\Delta_L} v_L^2
	+ a_1 v_L^2 v_R^2
	+ 2 ( a_2 v_u^2 + a_3 v_d^2 ) v_R v_L + \cdots
\eea
which gives, after extremizing the potential with
respect to $v_L$,
\beq
\label{v3}
v_L = \frac{( a_2 v_u^2 + a_3 v_d^2 ) v_R }
	{M^2_{\Delta_L} + a_1 v_R^2}
\eeq

The minimal assumption one can make for
the fermion masses is to take just one complex
10-dimensional Higgs representation.
In this case one would have $Tr (m_u) = a v_u + b v_d$
while $Tr (m_d) = - a v_d - b v_u$ where the VEVS of
the two standard model doublets in the 10-plet are
denoted by $v_u$ and $v_d$ and $a$ and $b$ are constants.
In order to have $m_b \ll m_t$ one must assume $b \ll a$.

One may forbid the $b$-term either with a global continuous
Peccei-Quinn symmetry or via a discrete symmetry D.
In the first case if the Peccei-Quinn
symmetry is broken softly at the weak scale that would
lead to the existence of many higgs bosons contributing
to the up and down fermion masses at low energies. This
in turn would destroy natural flavour conservation
in weak interactions. For this reason we rule out this
possibility and assume instead the existence of a
discrete symmetry D, defined by tables 1 and 2.

We use in total three complex SO(10) 10-plets
two transforming as 5-plets and one as a singlet under the SU(2)$_H$
symmetry and, in addition, two 126 representations. The (126,5)-plet
is assumed to have a large positive $(\rm mass)^2$ of order the GUT
scale, so that it does not break the B-L symmetry and therefore
does not contribute to the Majorana mass $M_R$.

It is easy to see that, owing to our discrete symmetry D,
there is no mixing at the tree level between higgs doublets
that contribute to up and down quark mass matrices and we
can therefore assume that in the low energy theory there
is only one pair of such doublets, one coupled to up
and the other to down fermions. Alternatively, if there
is only one fine tuning in the theory (corresponding to
the usual hierarchy problem) then the down-type fermions
such as the $b$ quark can naturally get masses due to higgs
boson mixing, after radiative corrections, as illustrated in
Fig. 1 \cite{SO10}.

The corresponding $SO(10) \ot SU(2)_H \ot D$
invariant Yukawa couplings have the form
\beq
\label{yukawa}
\L = (16,3)
	\left[ \lambda_1 (10,5)_1^* +
	\lambda_2 (10,5)_2 +
	\lambda_3 (10,1)^* +
	\lambda_4 (\overline{126},5) +
	\lambda_5 (\overline{126},1) \right] (16,3) + h.c.
\eeq
where all fields are given in tables 1 and 2.

After electroweak breaking the fermions acquire
Dirac mass matrices which may be written as
\bea
M_{u,ab} = \lambda_1 u_{1ab}+ \lambda_2 u_0 \delta_{ab}
+ \lambda_3 u_{2ab} + \lambda_4 {\omega_u}_{ab}\\
M_{d,ab} = \lambda_1 d_{1ab}+ \lambda_2 d_0 \delta_{ab}
+ \lambda_3 d_{2ab} + \lambda_4 {\omega_d}_{ab}\\
M_{\ell,ab} = \lambda_1 d_{1ab}+ \lambda_2 d_0 \delta_{ab}
+ \lambda_3 d_{2ab} - 3 \lambda_4 {\omega_d}_{ab}\\
M_{D,ab} = \lambda_1 u_{1ab}+ \lambda_2 u_0 \delta_{ab}
+ \lambda_3 u_{2ab} -3 \lambda_4 {\omega_u}_{ab}
\eea
where $u_0$ and $d_0$ are the VEVS of
the (10,1) along the $5$ and $\bar{5}$
and the other VEVS are defined as
\bea
u_{1ab} = \VEV{(10,5)_1} \: \:
u_{2ab} = \VEV{(10,5)_2} \: \:
\eea
while ${\omega_u}_{ab}$ is the VEV of the colour
singlet part of the (15,2,2) of (126,5).

Note that all quark masses and mixings, as well as
charged lepton masses can be correctly reproduced.
In addition the induced VEV of the (126,5) plays an
important role in giving an acceptable pattern
of Dirac mass matrices for the fermions. On the other
hand the (126,1) acquires a large VEV $v_R$ which violates
B-L and gives a large mass $M_R$ to \rh \neus proportional
to $\lambda_5$, and thus to the identity matrix.

In the case where there is only one light higgs
doublet, one may write
\bea
M_{u, ab} = S_{ab} + \frac{m_b}{m_t} S_{1ab}\\
M_{d, ab} = S_{ab} \frac{m_b}{m_t} + S_{1ab}\\
M_{\ell,ab} = S_{ab} \frac{m_b}{m_t} + S_{2ab}\\
M_{D,ab} = S_{ab} + \frac{m_b}{m_t} S_{2ab}
\eea
where $Tr(S)=m_t$, $Tr(S_1)=Tr(S_2)=0$.

Let us now turn to the discussion of the seesaw formula for neutrino
masses, obtained from our model. Since $v_R = \VEV{\Delta_R}$ breaks
the B-L symmetry, it gives rise to the heavy Majorana mass of
the right-handed neutrinos. The modified $\nu_L$-$\nu_R$
seesaw mass matrix takes the form
\beq
\label{seesaw1}
\left(\begin{array}{ccccc}
	\lambda v_L & M_D \\
	M_D^T & \lambda v_R
\end{array}\right)
\eeq
where $\lambda$ and $M_D$ are $3\times3$ matrices
and $v_L$ is given in \eq{v3}. Since $v_R$ arises from
an SU(2)$_H$ singlet it follows that the $\lambda$ matrix
in \eq{seesaw1} is proportional to the identity matrix,
since $\lambda \equiv \lambda_5$. The light neutrino
mass matrix that follows from diagonalizing \eq{seesaw1} is
given as
\beq
\label{mnu}
m_\nu \approx \lambda v_L - ({\lambda v_R})^{-1} M_{D} M^T_{D}
\eeq

Using the two-loop renormalization group equations
and last measurements of the electroweak mixing parameter
$\sin^2 \theta_W$ and the strong coupling constant $\alpha_s$
derived from LEP and other data we determine the mass of the
\rh bosons and, as a result, find
$v_R \simeq 0.8 - 1.1 \times 10^{14}$ GeV.
Taking $\lambda \simeq 1.5$, $a_1 \simeq .2$
(note that $M_{\Delta_L}^2 \sim M_{\Delta_R}^2 \simeq .3 v_R^2$)
$a_2 \simeq 1$ in \eq{v3} we get for the \neu masses
$m_{\nu_e} \simeq m_{\nu_\mu} \simeq m_{\nu_\tau} \simeq 2$ eV,
the scale needed for the hot dark matter.
On the other hand we find for the \ne - \nm
and \nm - \nt mass splittings
\bea
\Delta m_{e \mu} = \kappa_c \frac{m_c^2}{\lambda v_R} \\
\Delta m_{\mu\tau} = \kappa_t \frac{m_t^2}{\lambda v_R}
\eea
where we have determined $\kappa_c \simeq .05-.07$
and $\kappa_t \simeq .2$ from the renormalization
group equations. Now using $m_c \simeq 1.5$ GeV
and $m_t \simeq 160 $ GeV one finds just the
right narrow range allowed for the MSW conversions.
For the atmospheric \neu oscillations we find
$\Delta m_{\mu\tau} \simeq 4.5 \times 10^{-2}$ eV,
corresponding to 0.18 eV$^2$ for the squared mass difference.
Taking into account uncertainties such as in the
charm and top masses, one sees that the solution
is quite consistently obtained in our model.
Note that the induced triplet VEV $v_L$ gives the main
contribution to the light \neu masses $m = \lambda v_L$.
and, to a good approximation, the Dirac neutrino mass
matrix {\sl is determined by the up quark masses, leading to
the successful prediction \eq{ratio} for the required mass
splittings in \eq{sol0} and \eq{atm0}. }

These mass differences are just right to
explain the solar neutrino deficit and the
atmospheric neutrino anomalies, while the
absolute light \neu mass scale 2 eV roughly
gives the value required for \neus to provide
the hot dark matter.

The 12 mixing angle is chosen in the range
of \eq{sol0} while the 23 angle is chosen to
fit the atmospheric \neu deficit. This angle
is quite large and, in our SO(10) model it
arises from the diagonalization of the charged
lepton mass matrices, and is therefore unrelated with the
corresponding angle of the Kobayashi-Maskawa
matrix of the quark sector. The 13 angle is
basically free and may give rise, in a three
parameter fit of the data, to somewhat wider
allowed ranges for our solution than given by
\eq{sol0} and \eq{atm0}.

On the other hand, as illustrated in Fig.2, the
spontaneous breaking of the
global SU(2)$_H$ family symmetry with SO(10)
singlet, SU(2)$_H$ doublets higgs fields
can be achieved in such a way that there
is no mixing between the (126,1) and (126,5)
multiplets. As a result, the degenerate \neu
spectrum prediction is unaffected.

\section{Discussion}

We have proposed a simple SO(10) scheme where
quasi-degenerate 2 eV \neu masses naturally arise.
The degenerate $\nu_e$, $\nu_\mu$ and
$\nu_\tau$ \neus fit the hot dark matter inferred
from large scale structure simulations. On the
other hand the splittings between \ne and \nm
explain the solar \neu deficit and that between
\nm and \nt fits well the atmospheric \neu anomaly.
Therefore all present hints in favour of massive
\neus are nicely reconciled. The 2 eV mass of the
majorana neutrinos can lead to a neutrinoless
double beta decay rate that could be observable
in enriched germanium experiments.

The model is based on a realization of
the seesaw mechanism in which the leading
contribution to the light \neu masses is universal,
due to a suitable SU(2) horizontal symmetry, broken
only in the other sectors of the theory. The model
is in good agreement with all relevant observations
such as the proton decay limits and the measurements
of $\alpha_s$ and $\sin^2  \theta_W$ that
come from LEP and other experiments.

{\bf Acknowledgements}

This work was supported by DGICYT under grant number
PB92-0084, by the EEC grant CHRX-CT93-0132, and by the
Univ. of Valencia-Yerphi exchange program (A.I.).
We thank A. Joshipura for discussions and for bringing
his paper \cite{joshi} to our attention at the WHEPP-3
workshop held in Madras, January, 1994.

\vfill

Note added: As we finished this work we became aware
of two recent papers described in ref. \cite{new}.
The second deals mostly with the phenomenology of
degenerate \neus at the \321 level, while the first
also gives an SO(10) realization.

\newpage

{\bf Figure captions}\\

{\bf Fig.1} \\
Typical graph contributing to the mixing between up
and down-type higgs bosons\\

{\bf Fig.2} \\
Typical graphs breaking the SU(2) family symmetry
in the higgs sector

\newpage
\begin{table}
\begin{center}
\caption{Quantum Numbers and VEVS of Higgs Bosons}
\begin{displaymath}
\begin{array}{|c|c|c|}
\hline
\mbox{$SO(10) \ot SU(2)_H$} & \mbox{$SU(4) \ot SU(2)_L \ot SU(2)_R$}
& \mbox{$D$} \\
\hline
(54,1) & (1,1,1) & 1 \\
(45,1) & (15,1,1) & -1 \\
(\overline{126},1) & (10,1,3) + (\overline{10},3,1) & i \\
(\overline{126},5) & (15,2,2) & i \\
(10,5)_1 & (1,2,2) & -i \\
(10,5)_2 & (1,2,2) & i \\
(10,1) & (1,2,2) & -i \\
\hline
\end{array}
\end{displaymath}
\end{center}
\end{table}
\begin{table}
\begin{center}
\caption{Quantum Numbers of Matter Fields }
\begin{displaymath}
\begin{array}{|c|c|}
\hline
\mbox{$SO(10) \ot SU(2)_H$} & \mbox{$D$} \\
\hline
(16,3) & \exp{-i\pi/4} \\
\hline
\end{array}
\end{displaymath}
\end{center}
\end{table}

\end{document}